\documentclass{FBSsuppl}
\usepackage{amsfonts}
\usepackage{amssymb}

\title{An Effective Field Theory for the Three-Body System}
\author{T. Barford, M. C. Birse}
\institute{Theoretical Physics, Department of Physics and Astronomy, University of Manchester, Manchester, M13 9PL, UK.}

\begin{document}

\maketitle
\begin{abstract}
We study the scattering of a particle from a bound pair in an effective field
theory using a distorted-wave renormalisation group method
to find the power-counting for the three-body force terms. We find that 
three-body terms appear at lower orders than naively expected. 
They start with a marginal term that varies as a logarithm rather than a power 
of the energy scales in the problem. The marginal term has 
important implications for the three-body problem in nuclear physics.
\end{abstract}

\section{Introduction}

The need for a model-independent approach to the three-nucleon system is clear.
The plethora of two-body potentials, each capable of describing two-body data 
extremely well, fail when applied to the three-body problem without an 
additional three-body force constructed to fit the data.

Effective Field Theories (EFTs) offer a systematic and model-independent 
treatment of nuclear and hadronic physics at low energies \cite{rev1}. In an EFT we hope 
to take advantage of a separation of scales by expanding observables in 
powers of the ratio $Q/\Lambda_0$, where $Q$ denotes a generic low-energy 
scale and $\Lambda_0$ a typical scale of the underlying physics. The 
expansion will be useful if the seperation of scales is large enough to 
ensure quick convergence.

An EFT is defined by a Lagrangian containing all possible local terms 
consistent with the symmetries of the underlying theory. Although this will 
invariably lead to an infinite number of terms, we hope to organise them 
according to some "power-counting", related to the number of low-energy 
scales in each term. EFTs are well-developed and understood for two-body
physics and now the three-body problem is an area of keen interest.

The mathematical tool that allows us to determine the power-counting is the 
renormalisation group (RG) \cite{bmr}. If we are concerned with problems 
where the 
wavelengths of the particles are far longer than the range of the 
interaction, the scattering is insensitive to the detailed structure of that 
interaction. Replacing the interaction with contact interactions and forming 
an EFT Lagrangian leads to UV divergences from the high-momentum 
modes of loop diagrams. Accordingly we regulate the loop diagrams with a 
floating UV cut-off $\Lambda$. Since all observables should be independent 
of this cut-off we renormalise the theory by absorbing the resulting 
$\Lambda$-dependence into the EFT couplings. Solving the resulting RG 
equation provides a frame-work for constructing power-counting schemes. The 
resulting EFT absorbs the effects of the short-range physics and 
parameterises it on a power-series in $Q$.

The key to constructing power-counting schemes is fixed points. After rescaling all quantities in terms of $\Lambda$, we may look for solutions to the RG equation which are independent of $\Lambda$. Such solutions are referred to as fixed points. Since these have no scale attached to them solutions of the RG equation tend to one of them as $\Lambda\rightarrow 0$. Study of the scaling behaviour of solutions close to the fixed points leads to different power-counting schemes.

\section{Two-body EFT}

The simplest EFT has only one field, the field of the 
asymptotic particles being considered. Such a theory is known as a 
``pionless'' EFT in nuclear physics. For a two-body system the Lagrangian 
takes the form,
\begin{equation}
{\cal L}_{2B}=\psi^{\dagger}\left(i\partial_0+\frac{\nabla^2}{2M}\right)
+\frac{C_0}{4}|\psi|^4
+\frac{C_2}{4}\left(\psi^\dagger(\overrightarrow\nabla
-\overleftarrow\nabla)^2\psi^\dagger\right)\psi^2+{\rm H.c.})
+\ldots
\end{equation}
Using the RG method outlined above one may construct two different 
power-counting schemes for such a system. The RG 
equation for the leading term $C_0$ is,
\begin{equation}
\Lambda\frac{\partial\hat C_0}{\partial\Lambda}=\hat C_0(1+\hat C_0),
\end{equation}
where the hat signifies a dimensionless rescaled coupling. There are clearly two 
fixed point solutions to this equation, $\hat C_0=0$ and $\hat C_0=-1$. These 
lead to the two different schemes.

Perturbing about the fixed point $\hat C_0=0$ leads to Weinberg or naive 
counting in which the leading perturbation scales with $Q$. This counting 
system is useful for weakly interacting systems.

Perturbing about the second fixed point $\hat C_0=-1$ leads to KSW counting, 
in which the leading perturbation scales with $Q^{-1}$. The fixed-point 
solution corresponds to a system with a bound-state at exactly zero energy, 
consequently this expansion is useful for strongly attractive systems with 
a shallow bound-state and is the expansion most useful in nuclear and atomic 
physics. The counting scheme is equivalent to the effective range expansion.
The  couplings, $C_0$ and $C_2$, can be directly related to the scattering 
length, $a_2$, and effective range, $r_2$, respectively.

Once a power-counting scheme is chosen the free parameters in the couplings 
can be fixed using experimental data. The resulting theory is then predictive 
for other processes.

\section{Building a Three-Body EFT}

Our aim is to extend the KSW EFT to describe three-body interactions,
 by using the already determined two-body couplings and constructing a 
counting scheme for the three-body couplings,
\begin{equation}
{\cal L}_{3B}=\frac{D_0}{36}|\psi|^6+\frac{D_2}{36}
\left(\psi^\dagger(\overrightarrow\nabla-
\overleftarrow\nabla)^2\psi^\dagger\right)\psi^\dagger\psi^3+{\rm H.c.})
+\ldots
\end{equation}

In scattering of a third particle from a bound pair the two-body contact potential leads to a 
particle exchange force with a range $\sim a_2$. In nuclear physics 
in particular, the length scale set by the scattering length is far longer 
than any other scale in the problem. We hope to use this seperation of 
scales to construct an EFT for three-bodies.

We may treat the long-range physics, $Q\sim 1/a_2$ exactly by working in 
terms of the distorted waves (DWs) of the particle-exchange 
potential \cite{bb}. The tool for constructing the power-counting schemes is then the 
distorted wave renormalisation group (DWRG) in which the cut-off $\Lambda$ is 
applied in the basis of the distorted waves. This method ensures that all 
non-analytic behaviour resulting from the long-range 
potential is factored out in terms constructed from the DWs.

To apply the DWRG method to the three-body problem we must find the three-body
wavefunction for a system with a finite two-body
scattering length and zero effective-range. This two-body interaction can be 
expressed simply as a boundary condition on the wavefunction at zero 
seperation. For the three-body wavefunction this takes the form,
\begin{equation}
\left[\frac{\partial\psi(r_{23},r_{1})}{\partial r_{23}}
\right]_{r_{23}=0}
=-\frac{1}{a_2}\Bigl[\psi(r_{23},r_{1})\Bigr]_{r_{23}=0},
\end{equation}
where we are using the three-body Jacobian coordinate system. When this is
applied to the integral equation for the three-body wavefunction we obtain an 
integro-differential equation for the projection of the wavefunction, 
$\phi(r_1)=\psi(r_{23}=0,r_1)$. This is equivalent to the 
equation given by Skorniakov and Ter-Martirosian \cite{stm}. 
Generally, the equations for $\phi$ and $\psi$ must be solved numerically. 
In the zero-energy limit we can solve them
analytically. These analytic solutions give the short-range asymptotic 
behaviour of the full solutions
\begin{equation}
\psi(r_{23},r_{1})={\cal N}(ka_2)f(\theta)\sin(s_0 \ln(kR)+\eta(R_0)),
\end{equation}
where $R$ and $\theta$ are the hyperradius and hyperangle respectively, $f$ is
 some known function, $k=\sqrt{4(a_2^2+ME)/3}$, $s_0\approx 1$, $\cal N$
 is some numerically computed normalisation. Here, $\eta$ is an arbitrary 
phase that must be fixed by applying a boundary condition at some $R_0$, 
$\eta(R_0)=-s_0\ln(kR_0)$. This solution is precisely that found by Efimov 
\cite{efm}.

Using the DWRG method, we may use this result to construct the power-counting 
for the three-body system. One finds that the DWRG equation has no true fixed
point but a logarithmic evolution. Perturbations 
around this lead to an expansion of the form 
${\cal D}_{n,m}p^{2n}a_2^{-m}\Lambda_0^{-2n-m}$, where $p$ is the 
on-shell momentum, $n,m\ge 0$, ${\cal D}_{n,m}$ are free dimensionless 
parameters that should be of order one and $\Lambda_0$ is the scale at which 
the EFT breaks down. This expansion provides the power-counting for the 
system. 

The leading term is marginal, i.e. it is dimensionless and does not 
scale with $Q/\Lambda_0$. As expected, such a marginal 
perturbation leads to logarithmic behaviour in $Q$ which 
necessitates the introduction of a new scale. In this case the scale is in
fact $R_0$, the scale appearing in the boundary 
condition on the three-body wavefunction.

It is important to notice that since the leading three-body term is marginal,
it occurs at next-to-leading order in the expansion (two-body leading term 
occurs at $Q^{-1}$) rather than the naively expected next-to-next-to-leading 
order. 

The final result may be expressed as a DW
effective range expansion,
\begin{equation}
{\cal N}(pa_2)\cot(\delta_{3B})=-{\cal M}(pa_2)
+\sum_{n,m}{\cal D}_{n,m}p^{2n}a_2^{-m}\Lambda_0^{-2n-2m}.
\end{equation}
$\delta_{3B}$ is the correction to the phaseshift due to the three-body force 
and $\cal N$ and $\cal M$ are numerically computed functions.

\section{Conclusion}

This analysis shows the importance of the three-body force in this
system.
The marginal behaviour of this force means that it must be included
in the low-energy EFT. It, or equivalently the boundary condition on the 
three-body wave function at short distances, is required to resolve the
ambiguities 
in the three-body system with zero-range two-body forces.
In nuclear physics, its role can be seen from the fact that different
phenomenological
nucleon-nucleon potentials lead to quite different predictions for
three-nucleon
binding energies and scattering lengths. However, these predictions all
lie on a 
single ``Phillips line" \cite{phi} showing that their
differences can be expressed in 
a single effective parameter $R_0$, the scale appearing the marginal term
of the 
three-body potential.


\begin{thebibliography}{99}

\bibitem{rev1} S. R. Beane, P. F. Bedaque, W. C. Haxton, D. R. Phillips and M. J. Savage, nucl-th/0008064; P. F. Bedaque and U. van Kolk, nucl-th/0203055.
\bibitem{bmr} M. C. Birse, J. A. McGovern and K. G. Richardson, Phys. Lett. \textbf{B464} (1999) 169, [hep-ph/9807302].
\bibitem{bb} T. Barford and M. C. Birse, hep-ph/0206146.
\bibitem{stm} G. V. Skorniakov and K. A. Ter-Martirosian, Sov. Phys. JETP
\textbf{4} (1957) 648.
\bibitem{efm} V. N. Efimov, Sov. J. Nucl. Phys. \textbf{12}, (1971) 589. 
\bibitem{phi} A. C. Phillips, Nucl. Phys.\textbf{A107}, 1968 209.

\end{thebibliography}
\end{document}